\documentclass[
aps,
pre,
twocolumn,
showpacs,
superscriptaddress
]{revtex4-2}

\usepackage{graphicx}
\usepackage{dcolumn}
\usepackage{bm}
\usepackage{amsmath,amssymb,mathtools}
\usepackage{color}
\usepackage{appendix}

\definecolor{darkblue}{rgb}{0,0,0.6}
\definecolor{darkred}{rgb}{0.6,0,0}
\definecolor{darkgreen}{rgb}{0,0.6,0}

\usepackage[colorlinks=true,urlcolor=darkblue,citecolor=darkblue,linkcolor=darkred,hyperfootnotes=false]{hyperref}

\usepackage{soul}
\usepackage{multirow}
\usepackage{tikz}
\usepackage{hhline}
\usepackage{algorithm}
\usepackage{algorithmic}

\usepackage[para,online,flushleft]{threeparttable}
\usepackage{booktabs}
\usepackage{multirow}

\usepackage{verbatim}

\usepackage{appendix}

\usepackage{setspace}

\makeatother

\begin{document}

\title{Quantifying information flow along a stochastic trajectory}

\author{Yongjae Oh}
\affiliation{Department of Physics and Astronomy \& Center for Theoretical Physics, Seoul National University, Seoul 08826, Republic of Korea}

\author{Euijoon Kwon}
\affiliation{Quantum Universe Center, Korea Institute for Advanced Study, Seoul 02455, Republic of Korea}

\author{Yongjoo Baek}
\email{y.baek@snu.ac.kr}
\affiliation{Department of Physics and Astronomy \& Center for Theoretical Physics, Seoul National University, Seoul 08826, Republic of Korea}

\date{\today}
 
\begin{abstract}
    Stochastic information flow (SIF) quantifies information flow at the trajectory level, overcoming the limitations of conventional symmetric, ensemble-averaged measures. However, computational difficulties have hindered the empirical application of the SIF. In this work, we propose a scalable deep-learning method for estimating the SIF from general time-series data. Its applications to an exactly solvable two-particle model, Kuramoto oscillators, and empirical trajectories of interacting motile cells demonstrate the utility of SIF as a data-driven indicator of cooperative structures.
\end{abstract}


\maketitle

Previous decades have seen the emergence of {\em stochastic thermodynamics} as a theoretical framework for the thermodynamics of small systems~\cite{Sekimoto1988langevin,Seifert2012stochastic}. In such systems, we should redefine thermodynamic quantities such as heat, work, and entropy production as trajectory functionals, whose values fluctuate due to microscopic noise. Theoretical advances of the field have identified a plethora of fluctuation theorems that govern the statistics of those thermodynamic quantities~\cite{Evans1993probability,Jarzynski1997,crooks1999entropy, Seifert2005entropy, kwon2011nonequilibrium, noh2012fluctuation}, and experimental advances have provided access to empirical measurement of the quantities in various small systems, such as biomolecular systems~\cite{Collin2005verification, Hayashi2010fluctuation}, artificial colloidal particles~\cite{lee2015nonequilibrium, Martínez2016brownian, wang2002experimental, wang2005experimental}, and electric circuits~\cite{van2004power, garnier2005nonequilibrium, falcon2009fluctuations, granger2015fluctuation}.

Meanwhile, recent advances have also incorporated another important quantity, {\em information}, into the realm of thermodynamics. Studies have revealed various forms of inequalities that modify the second law of thermodynamics using the information exchange, which resolve the paradox of {\em Maxwell's demon}~\cite{sagawa2008second, sagawa2010generalized, sagawa2012fluctuation, sagawa2012nonequilibrium, paneru2020efficiency}. In particular, formulations of Maxwell's demon as an externally imposed measurement--feedback protocol revealed {\em mutual information} as a key indicator of information exchange. Mathematically, the mutual information between two subsystems $X$ and $Y$ at time $t$ is defined as
\begin{align} \label{eq:mutual_info}
	I(X_t,Y_t) = H_{X_t} + H_{Y_t} - H_{X_t,Y_t}
	\;,
\end{align}
where $H_X$ denotes the Shannon entropy of the random variable $X$. A broad range of empirical studies, ranging from biomolecular signaling pathways and cellular processes~\cite{Tkačik2008information,Uda2013robustness,Dubuis2013positional} to interacting neurons~\cite{Barzon2025excitation},
have utilized mutual information to characterize the information exchange. However, being symmetric under the exchange of subsystems and lacking explicit dependence on their dynamics, mutual information is not a suitable measure of directed information flow.

To overcome this drawback, other studies~\cite{Allahverdyan2009thermodynamic,Horowitz2014thermodynamics,Hartich2014stochastic,horowitz2014second,Horowitz2015multipartite, Parrondo2015thermodynamics} have introduced a quantity called {\em information flow} (IF). Its motivation follows from the decomposition $d_t I = \dot{I}^X + \dot{I}^Y$, where 
$\dot{I}^X = \partial_{t'} I(X_{t+t'},\, Y_t) \big|_{t' \rightarrow 0}$
is the IF from $Y$ to $X$ at time $t$, with $\dot{I}^Y$ similarly defined. This quantity is explicitly dependent on the dynamics and asymmetric under the exchange of $X$ and $Y$, with $\dot{I}^X > 0$ ($\dot{I}^X < 0$) indicating that $X$ gains (loses) information about $Y$ by the change $X_t \rightarrow X_{t+dt}$. This formalism is useful for describing Maxwell's demon in autonomous dynamical systems~\cite{Horowitz2015multipartite, amano2022insights, penocchio2022information, tanogami2023universal, leighton2023inferring, leighton2024information}, in which measurement and feedback occurs entirely via internal dynamics.

Just as stochastic thermodynamics redefines thermodynamic quantities as trajectory-level variables, it is possible to formulate a stochastic version of the IF, which has been called {\em stochastic information flow} (SIF)~\cite{Rosinberg2016continuous}. The SIF quantifies the trajectory-wise structure of information dynamics, which is especially important when subsystems are identical. In such cases, whereas the IF is zero in the steady state, the SIF can still characterize the fluctuating information exchange when the subsystems alternate the roles of information source and recipient. However, previous studies of the SIF have focused only on its applications to universal thermodynamic relations, such as fluctuation theorems and generalized second-law inequalities~\cite{Rosinberg2016continuous,Shiraishi2015fluctuation,Gopal2024information}. In this Letter, we study the SIF statistics in concrete physical systems and explore its utility as an empirical, trajectory-level indicator of information exchange and Maxwell's demon.

\emph{Identification of stochastic information flow.} --- Let us consider a system consisting of two mutually interacting subsystems $X$ and $Y$. The system evolves according to a Markov process, whose state at time $t$ is $(X_t, Y_t)$. For every possible instance $(x,y)$ of the system's state, one can obtain the pointwise mutual information (PMI)
\begin{align}
    i_{X_t,Y_t} (x,y) = \ln \frac{p_{X_t, Y_t}(x,y)}{p_{X_t}\!(x)\,p_{Y_t}\!(y)}
    \;,
\end{align}
where $p_{X_t}$ and $p_{Y_t}$ are marginal distributions of the joint probability density function $p_{X_t, Y_t}$ of the system's state. While the PMI is a value assigned to a certain state $(x,y)$, one can also define the stochastic mutual information (SMI) $\mathcal{I}_t (X_t, Y_t) = i_{X_t,Y_t} (X_t,Y_t)$~\cite{Parrondo2015thermodynamics,Dabelow2019irreversibility,Takaki2022information,Nicoletti2024tuning}, which is a random variable whose value is yet to be fixed by a probabilistic instance of $(X_t, Y_t)$. Mirroring Eq.~\eqref{eq:mutual_info}, the SMI satisfies
\begin{align} \label{eq:SMI_SI}
\mathcal{I}_t (X_t, Y_t)= S_{X_t} (X_t)+ S_{Y_t} (Y_t) - S_{X_t, Y_t} (X_t, Y_t)
\;,
\end{align}
where $S_{X_t}\!(x) = -\ln p_{X_t}\!(x)$ is the self-information (also called the stochastic entropy if $x$ is still random) of an instance $x$ of $X_t$, with $S_{Y_t}$ and $S_{X_t, Y_t}$ defined similarly. Clearly, the mutual information is the mean of the SMI, with Eq.~\eqref{eq:mutual_info} recovered by averaging Eq.~\eqref{eq:SMI_SI} side by side.

In general, the time derivative of the SMI satisfies
\begin{align} \label{eq:SMI_decomp}
d_t \mathcal{I}_t  = \dot{\mathcal{J}}_{X_t} + \dot{\mathcal{J}}_{Y_t} +  \partial_{t'} \mathcal{I}_{t'} (X_t, Y_t) |_{t' =t}
\;,
\end{align}
where we have identified the SIFs
\begin{align} \label{eq:SIF_definition}
\dot{\mathcal{J}}_{X_t} = \partial_{t'} \mathcal{I}_t(X_{t'}, Y_t)|_{t' =t},~
\dot{\mathcal{J}}_{Y_t} = \partial_{t'} \mathcal{I}_t(X_t, Y_{t'})|_{t' =t}
\;.
\end{align}
We note that the time derivative $\partial_{t'}$ in the SIFs applies to the state of a subsystem, while the same operator in the last term of Eq.~\eqref{eq:SMI_decomp} applies to $p_{X_{t'},Y_{t'}}$, $p_{X_{t'}}$, and $p_{Y_{t'}}$ composing the SMI. Thus, the last term of Eq.~\eqref{eq:SMI_decomp} vanishes in the steady state, allowing us to write $d_t \mathcal{I}_t  = \dot{\mathcal{J}}_{X_t} + \dot{\mathcal{J}}_{Y_t}$. When $\dot{\mathcal{J}}_{X_t}>0$ ($\dot{\mathcal{J}}_{X_t}<0$) for an infinitesimal trajectory, the change $X_t \rightarrow X_{t+dt}$ increases (decreases) the information about $Y$ held by $X$, which can be interpreted as $X$ learning (forgetting) about $Y$.

Although the formalism described so far are applicable to any Markovian bipartite system, we focus on the case where $(X_t, Y_t)$ evolves continuously according to the Langevin equations. For this case, an explicit formula for the SIF reads
\begin{align} \label{eq:SIF_rate}
    \dot{\mathcal{J}}_X = \nabla_{x} \ln p_{Y_t | X_t} (x,Y_t) \big|_{x=X_t} \circ \dot{X}_t\;,
\end{align}
where $\circ$ is a Stratonovich product, and $p_{Y_t|X_t}$ denotes the conditional probability distribution function of $Y_t$ given $X_t$. We can obtain the relation for $\dot{\mathcal{J}}_Y$ by exchanging $X_t$ and $Y_t$ in the above expression. See Appendices~\ref{sec:app_stoc_info_langevin} and \ref{sec:app_stoc_info_MJ} for a derivation of this formula as well as an analogous formula which holds for discrete Markov jump processes.

\emph{Maxwell's demon in a two-particle system.} --- As an exactly solvable model demonstrating the utility of the SIF as a marker of Maxwell's demon, we consider two one-dimensional overdamped Brownian particles in contact with a heat bath at temperature $T$. The particles are confined by identical harmonic traps of stiffness $k$ and mutually coupled by a spring of stiffness $K$. Their equations of motion read
\begin{align}
\label{eq:equil_example}
    \gamma\dot{X}_t &= -kX_t-K(X_t-Y_t)+\sqrt{2\gamma T}\,\xi_X\!(t)\;, \nonumber
    \\
    \gamma\dot{Y}_t &= -kY_t-K(Y_t-X_t)+\sqrt{2\gamma T}\,\xi_Y\!(t)\;,
\end{align}
where $\gamma$ is the friction coefficient, and $\xi_X$ and $\xi_Y$ are independent Gaussian white noises with unit intensity.

This system satisfies the fluctuation--dissipation relation (ensured by the thermal noise amplitude $\sqrt{2\gamma T}$) with only a single temperature $T$, so it eventually reaches thermal equilibrium.
Since we can obtain the exact equilibrium statistics of the system, it is easy to evaluate the SIF of each trajectory by integrating Eq.~\eqref{eq:SIF_rate}.
To this end, we simulate Eq.~\eqref{eq:equil_example} and evaluate the time-accumulated SIF
$\mathcal{J}_{X}(\tau)\equiv \int_{0}^{\tau}\dot{\mathcal{J}}_{X_t}dt$
over a time window $[0,\,\tau]$. We present how this SIF correlates with the stochastic area $\dot{A}_{XY}=-Y_t\circ\dot{X}_t+X_t\circ\dot{Y_t}$~\cite{Levy1940lemouvement}, which has been used as an empirical measure of irreversibility~\cite{Ghanta2017fluctuation,Gonzalez2019experimental,duBuisson2023large}. We also note that $A_{XY} > 0$ ($A_{XY} < 0$) is achieved when the oscillator $X$ precedes (follows) the oscillator $Y$ by some phase difference between $0$ and $\pi$, which means that $X$ ($Y$) moves away from $Y$ ($X$) as $Y$ ($X$) chases $X$ ($Y$), like a predator--prey pair. In this sense, $A_{XY}$ indicates the non-reciprocal roles played by the two particles.

\begin{figure}
	\includegraphics[width=0.48\textwidth]{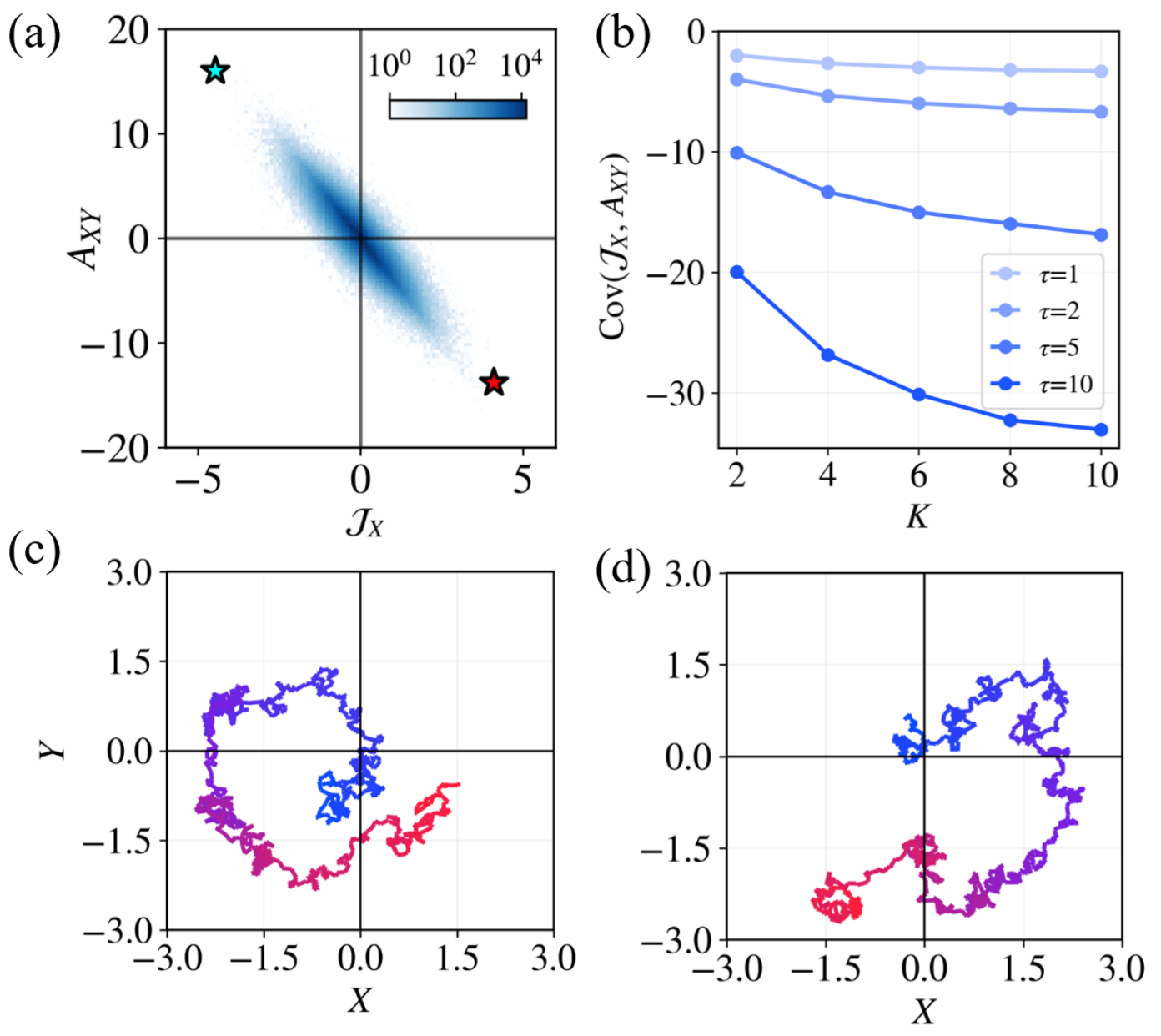}
    \caption{\label{fig:two_particle}
    (a) Histogram of time-integrated SIF $\mathcal{J}_X$ versus stochastic area $A_{XY}$ of the two-particle model, collected for $10^6$ trajectory segments with observation period $\tau=1$. The parameters used for the simulation are $k=2$, $K=2$, $\gamma=1$, $T=2$, and the discrete time step is fixed at $\Delta t = 0.001$. The color bar indicates the number of trajectories inside each bin. (b) Covariance between the SIF and the stochastic area measured for $10^5$ trajectories, as $\tau$ is varied. We also show the sample trajectories achieving the (c) minimum [$\mathcal{J}_X(\tau)=-4.472$] and the (d) maximum [$\mathcal{J}_X(\tau)=4.096$] values of $\mathcal{J}_X(\tau)$. The colors indicate the time evolution from the initial (blue) to the final state (red).}
\end{figure}

As demonstrated in Fig.~\ref{fig:two_particle}(a) by a heatmap showing the joint distribution of $\mathcal{J}_{X}$ and $A_{XY}$, the two variables are anticorrelated, with $|A_{XY}|$ tending to increase with $|\mathcal{J}_{X}|$. Moreover, for various values of the observation period $\tau$, the anticorrelation becomes stronger with the interpaticle coupling $K$. To visualize the dynamical effects of the SIF, in Fig.~\ref{fig:two_particle}(c, d) we illustrate the trajectories associated with the minimum and the maximum observed values of $\mathcal{J}_X(\tau)$ [also indicated by two stars in Fig.~\ref{fig:two_particle}(a)], respectively, with the flow of time indicated by the trajectory color changing from blue (initial state) to red (final state). The trajectories exhibit circular shapes in these examples, demonstrating that $X$ can stochastically erase information about $Y$ [$\mathcal{J}_X(\tau) < 0$] by trying to move away [see the counterclockwise motion in Fig.~\ref{fig:two_particle}(c)], while $X$ can also stochastically gain information about $Y$ [$\mathcal{J}_X(\tau) < 0$] by trying to chase it [see the clockwise motion in Fig.~\ref{fig:two_particle}(d)]. In the former (latter) case, $Y$ ($X$) plays the role of Maxwell's demon to $X$ ($Y$).

To sum up, even in an equilibrium system with identical subsystems where the ensemble-averaged IF is zero, a coordination reminiscent of Maxwell's demon can arise at the trajectory level. The SIF provides a principled way to uncover this stochastic structure by conditioning on its atypical values.

\begin{figure}
	\includegraphics[width=0.47\textwidth]{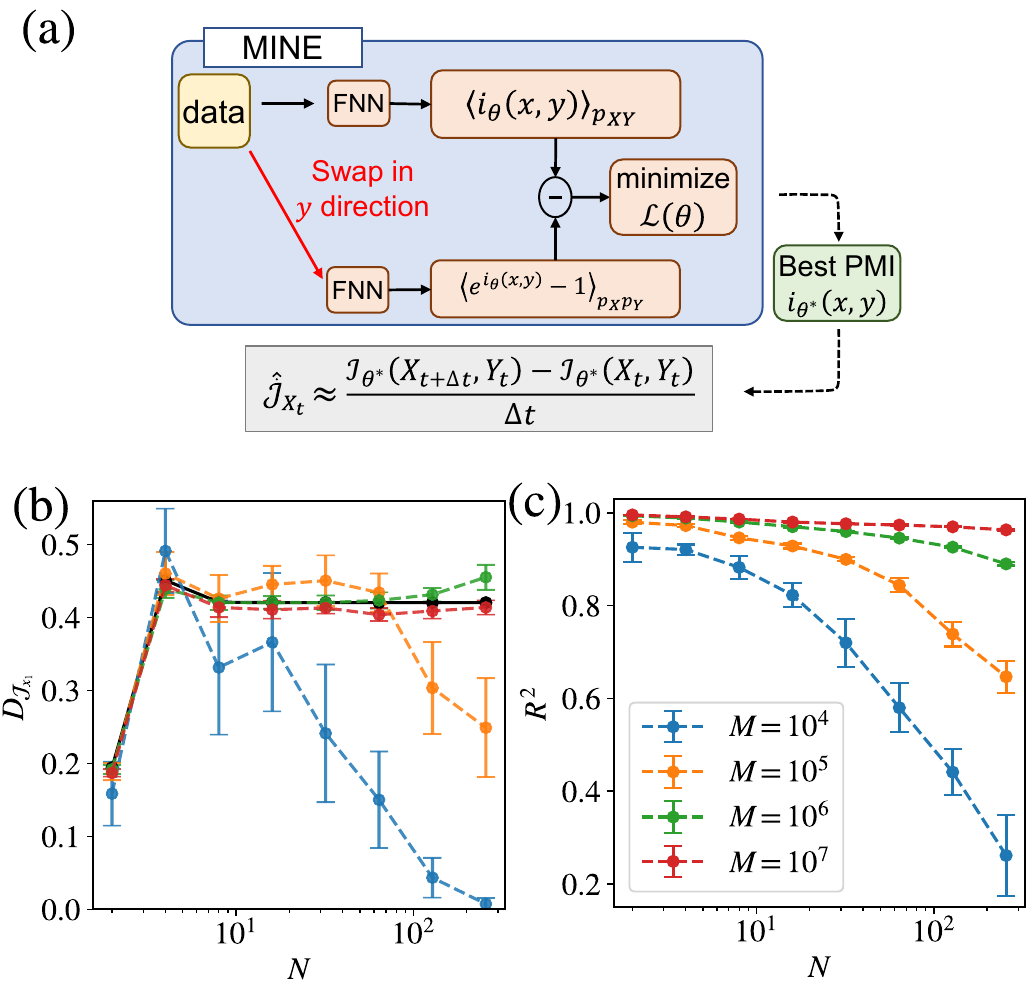}
    \caption{\label{fig:N_bead} (a) Schematic of NESIF. $\Delta t = 0.01$. (b, c) Test results of NESIF using the $N$-bead model with $k=K=\gamma=T=1$. (b) Comparison between the exact scaled variance $D_{\mathcal{J}_{x_1}}$ (solid line) and the values inferred using the NESIF (symbols), as the dataset size $M$ is varied. (c) The coefficient of determination $R^2$ between $\mathcal{J}_{x_1}(\tau = 1)$ and $\hat{\mathcal{J}}_{x_1 }(\tau = 1)$. In (b) and (c), the error bars denote standard deviation of 10 independent trainings, and the lines are to guide the eye.}
\end{figure}

\emph{Estimation of the SIF via neural networks.} --- For homogeneous Langevin systems with linear drifts, the SIF can be calculated exactly via Eq.~\eqref{eq:SIF_rate} thanks to the analytically tractable steady-state statistics~\cite{Oh2026maxwell}. However, when such statistics are not available, calculation of the SIF becomes highly challenging. In theory, one may gather sufficiently many samples to obtain a high-resolution histogram of the steady-state distribution, from which one can calculate the SIF. In practice, this procedure is prone to the curse of dimensionality, especially when the dynamics involves many degrees of freedom. For practical applications to empirical time-series data, it is crucial to develop a highly scalable method for estimating the SIF.

Toward this goal, we propose a \emph{neural estimator of stochastic information flow} (NESIF). The core idea of the method is to estimate the SMI from data using an artificial neural network, and then to calculate the SIF from the SMI via its definition shown in Eq.~\eqref{eq:SIF_definition}.
For SMI estimation, we adopt the Mutual Information Neural Estimator (MINE)~\cite{Belghazi2018mutual}, in which a neural network learns the functional relationship between the instance $(x,y)$ of the two random variables $(X,Y)$ and their PMI $i_{X,Y}(x,y)$ by utilizing the variational representation
\begin{align} \label{eq:var_MI}
   I(X,Y) \ge \langle i_\theta (x,y) \rangle  _{p_{X,Y}} - \langle e^{i_\theta (x,y)} -1 \rangle_{p_X p_Y}
    \;,
\end{align}
where $\langle\cdot\rangle_p$ denotes an average with respect to the distribution $p$. The parametrized function $i_\theta(x,y)$, where $\theta$ denotes the internal state of the neural network, is the PMI estimator. Since the above inequality becomes an equality if and only if $i_\theta(x,y) = i_{X,Y}(x,y)$, one can train a neural network to estimate the SMI by finding $\theta$ that maximizes the rhs of Eq.~\eqref{eq:var_MI}.
The reader is referred to Appendix~\ref{sec:app_alpha_mine} for more details about how Eq.~\eqref{eq:var_MI}, or its generalization using the $\alpha$-divergence, estimates the PMI.

Suppose that the rhs of Eq.~\eqref{eq:var_MI} reaches the maximum when $\theta = \theta^*$. Then, we can estimate the SIF via
\begin{align} \label{eq:SIF_pred}
    \hat{\dot{\mathcal{J}}}_{X_t} = \frac{i_{\theta ^*} (X_{t+\Delta t}, Y_t ) - i_{\theta ^*}(X_t, Y_t)}{\Delta t}
    \;,
\end{align}
where $\Delta t$ should be small enough (we use $\Delta t \le 0.01$ throughout this work) to provide a good approximation for Eq.~\eqref{eq:SIF_definition}. In Fig.~\ref{fig:N_bead}(a), we present a schematic of the method described so far.

To check the method's reliability and scalability, we apply the NESIF to an $N$-bead model, which consists of a harmonic chain of $N$ identical overdamped beads at thermal equilibrium. They obey the Langevin equations
\begin{align} \label{eq:nbead}
    \gamma \dot{X}_t^n = -kX_t^n -K(2X_t^n - X_t^{n-1} -X_t^{n+1} ) + \sqrt{2\gamma T} \xi _n(t)\;,
\end{align}
where the coefficients $\gamma,\,k,\,K$, and $T$ have the same meanings as those appearing in Eq.~\eqref{eq:equil_example}, and the index $n$ runs from $1$ to $N$ with the periodic boundary conditions $X_t^0 = X_t^N$ and $X_t^1 = X_t^{N+1}$. As a measure of information exchange, we focus on $\mathcal{J}_{X^1} (\tau) = \int_0 ^\tau dt \,\dot{\mathcal{J}}_{X^1}(t) $, which is the time-integrated SIF into the single bead $X_t^1$ from the rest of the system $(X_t^2, \cdots ,X_t^N)$. Since every bead is identical, no bead can be a pure information source or recipient, so the mean of $\mathcal{J}_{X^1} (\tau)$ must be zero; however, the scaled variance $D_{\mathcal{J}_{X^1}} \equiv \text{Var}(\mathcal{J}_{X^1}(\tau))/(2\tau)$ of the SIF would still be nonzero, indicating the fluctuating component of information exchange.

Thanks to the linearity of Eq.~\eqref{eq:nbead}, we can analytically calculate $D_{\mathcal{J}_{X^1}}$, which will be detailed in~\cite{Oh2026maxwell}. In Fig.~\ref{fig:N_bead}(b), as $N$ is varied, we compare the exact values of $D_{\mathcal{J}_{X^1}}$ (black solid line) with its estimated values (symbols with error bars) for different values of $M$, the number of infinitesimal trajectory fragments. For $M=10^4$, the estimation is reliable up to $4$ oscillators. If $M=10^7$, the estimation stays in good agreement with the exact value up to $N = 256$. These are also confirmed by the coefficient of determination $R^2$ between the estimated and the exact values of the time-integrated SIF, as shown in Fig.~\ref{fig:N_bead}(c). These observations demonstrate that NESIF is a reliable and scalable estimator of the SIF.


\begin{figure} 
	\includegraphics[width=0.47\textwidth]{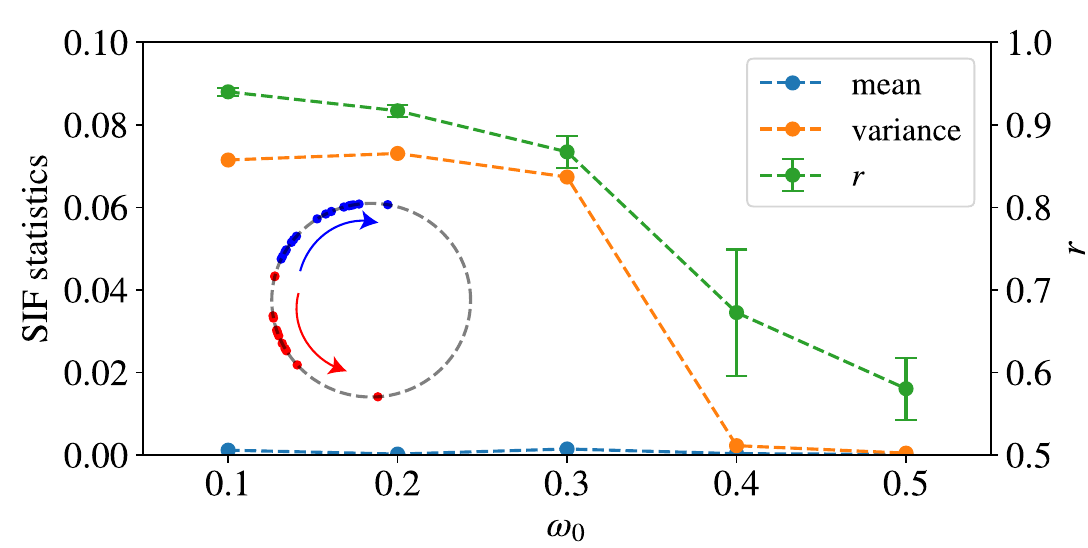}
    \caption{\label{fig:kuramoto}
    Mean (blue symbols) and variance (orange symbols) of the SIF from oscillators driven clockwise to those driven counterclockwise are plotted as functions of the driving frequency $\omega_0$, together with the synchronization order parameter $r$ (green). The trajectory data are obtained by simulating $32$ noisy Kuramoto oscillators with $K = 1$ using the time step size $\Delta t=0.01$ for the observation period $\tau = 10$. Inset: Schematic of the Kuramoto model with oppositely driven oscillators. The lines are to guide the eye.
    }
\end{figure}

\emph{SIF as a measure of cooperative effects.} --- Adopting NESIF, we explore whether the quantity can capture the existence of cooperation among identical elements of a system as it exhibits a macroscopic order. Here, we focus on the noisy Kuramoto model~\cite{Kuramoto1975selfentrainment,Sakaguchi1988cooperative,Gupta2014kuramoto,Imparato2015stochastic,Hong2020thermodynamic}. The model consists of $N$ coupled oscillators described by a system of overdamped Langevin equations given by
\begin{align}
    \dot{\theta}_n = \omega _n - \frac{K}{N} \sum_{m=1} ^N \sin (\theta_n - \theta_m ) + \sqrt{2T} \xi_n 
    \;,
\end{align}
where $K$ is the coupling strength, $T$ is the temperature of the heat bath, $\xi_n$ is an independent Gaussian white noise with unit intensity, and $\omega_n$ is the driving frequency of the $n$th oscillator sampled from the distribution $g(\omega)$ with the index $n$ running from $1$ to $N$. Although one cannot analytically calculate the steady-state distribution of this model, we know that the model exhibits a synchronization transition in the thermodynamic limit ($N\rightarrow \infty$), with an analytically computable critical temperature.

We demonstrate that the SIF successfully captures the strength of cooperative effects using a noisy Kuramoto system where a half of the oscillators are driven clockwise at frequency $\omega _0$, and the other half are driven counter-clockwise with the same frequency. Using the NESIF, we estimate the SIF from the oscillators driven counter-clockwise to the oscillators driven clockwise. Remarkably, even though the mean SIF always stays close to zero, the SIF variance reflects the phase transition of the system indicated by the order parameter $r$, as shown in Fig.~\ref{fig:kuramoto}. This result confirms the intuition that the two groups of oscillators actively exchange information only in the synchronized phase. We also apply the NESIF to the case where the system has no external driving and the synchronization is tuned by $K$, observing that the SIF reveals both cooperative effects and a characteristic time scale governing the fluctuations. See Appendix~\ref{sec:app_Kuramoto} for more detailed discussion.

\begin{figure}
	\includegraphics[width=0.5\textwidth]
    {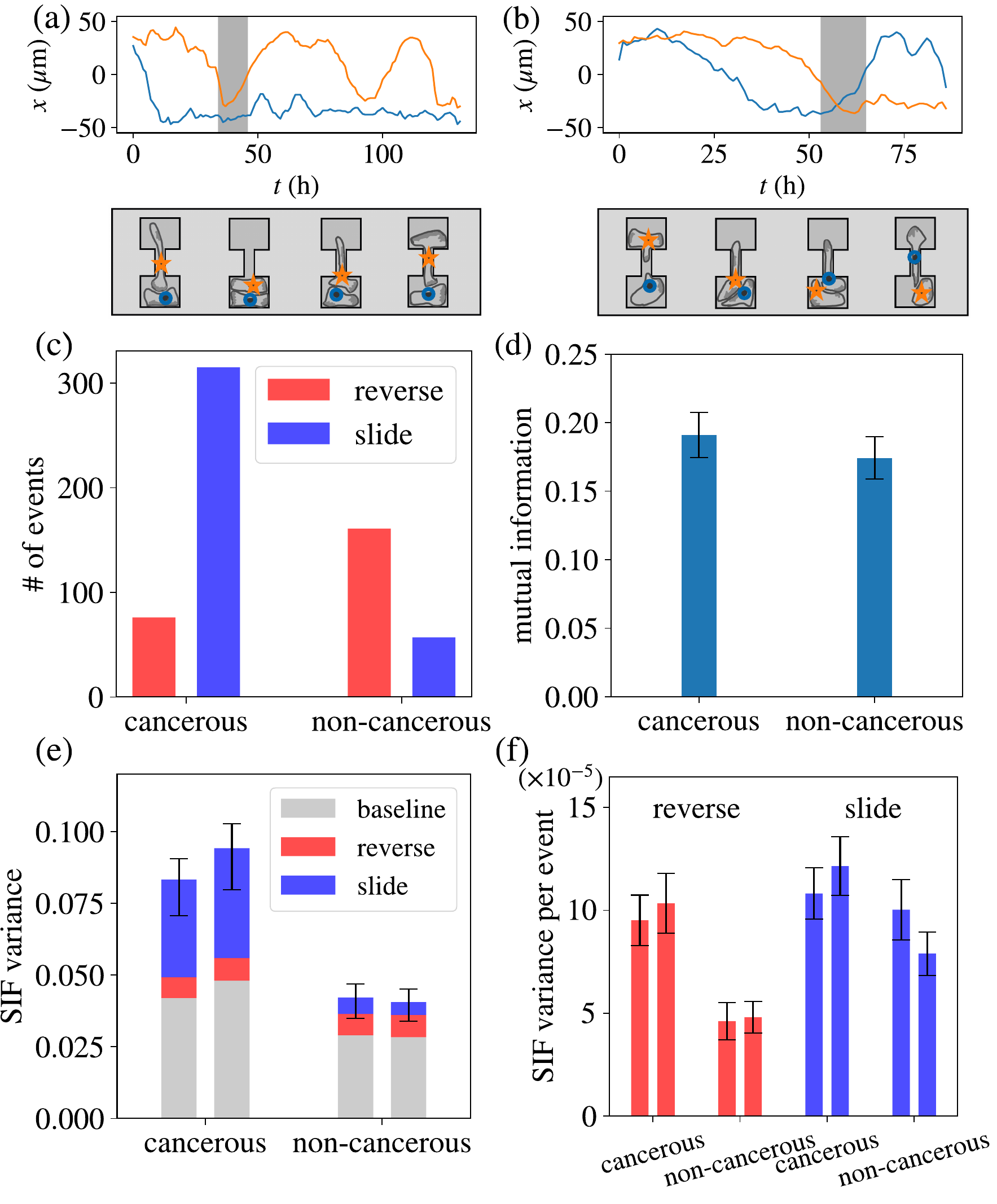}
    \caption{\label{fig:cell_cell} Sample trajectories and schematics of a pair of cells moving in an effectively one-dimensional channel, with (a) a reverse event and (b) a slide event highlighted by gray boxes. Data points are collected every one hour. (c) Frequencies of reverse and slide events and (d) mutual information between two interacting cells for each cell group. (e) SIF variance, with contributions from different types of collision events indicated by colors. (f) SIF variance per each collision event.
    }
\end{figure}

\emph{Application to two-cell dynamics.}---Finally, to demonstrate the applicability of the NESIF to empirical data, we use the method to quantify the information exchange between a pair of human mammary epithelial cells moving and interacting in an effectively one-dimensional channel~\cite{Brückner2021learning}, whose sample trajectories (indicating the positions of the cell nuclei along the channel) are shown in Fig.~\ref{fig:cell_cell}(a, b). When the cells collide, they either `reverse' [see Fig.~\ref{fig:cell_cell}(a)] so that their trajectories do not cross or `slide' [see Fig.~\ref{fig:cell_cell}(b)] so that their trajectories intersect each other. As shown in Fig.~\ref{fig:cell_cell}(c), cancerous cells (MDA-MB-231) tend to slide by each other, while non-cancerous cells (MCF10A) are more likely to reverse. Despite these differences, Fig.~\ref{fig:cell_cell}(d) shows that mutual information between the trajectories of the two cells fails to capture any significant difference between the two groups. In contrast, as shown in Fig.~\ref{fig:cell_cell}(e), the SIF variance is about twice larger for cancerous cells than for non-cancerous cells, revealing a clear difference between the two groups.

We investigate more detailed information exchange structure by dividing each cell trajectory into reverse, slide, and rest (or baseline) intervals [see the colors in Fig.~\ref{fig:cell_cell}(e)], which allows us to quantify the information exchange per each reverse or slide event, as shown in Fig.~\ref{fig:cell_cell}(f). For each slide event, the two cell groups show little difference in exchanged information. In contrast, for each reverse event, cancerous cells exchange much more information than non-cancerous counterparts. From these observations, we conclude that cancerous cells exchange more information due to (i) frequent slide events and (ii) vigorous information exchange for each reverse event. Although further biological implications of these behaviors are yet to be studied, our results demonstrate the utility of the SIF to empirical studies.

\emph{Conclusions.} --- Through this Letter, we have demonstrated the utility of stochastic information flow (SIF) as an indicator of information exchange and Maxwell's demon effects for both analytically tractable models and highly complex systems. For the latter, our work establishes a neural estimator of SIF (NESIF), a reliable and scalable data-driven method for studying the cooperative structures underlying collective behaviors beyond the mean level.

The NESIF is a highly versatile framework and can readily be applied to a great diversity of time-series data. It would certainly be interesting to apply the method to other kinds of dynamical systems---including chaotic systems~\cite{Staniek2008symbolic,Dickten2014identifying,Massaro2023on}, biological and artificial neural networks~\cite{Vicente2011transfer,Wibral2011transfer,Novelli2021inferring,Tishby2015deep,Koch-Janusz2018mutual}, biochemical networks~\cite{Tkačik2008information,Uda2013robustness,Dubuis2013positional,Moor2023dynamic,Imaizumi2022assessing}, climate and geophysical systems~\cite{Kleeman2007information,Hagan2019time,Tanogami2024information}, and socioeconomic systems~\cite{Marschinski2002analysing,Bossomaier2013information,BorgeHolthoefer2016dynamics,Tabatabaeian2025information}---and check whether there are new information exchange structures missed by ensemble-averaged quantities like mutual information and transfer entropy, as we explicitly illustrated by the cell dynamics. We also note that the NESIF assumes the system to be in the steady state; another important direction of future studies should be about extending the method to transient states.

\emph{Acknowledgements.}--E.K. and Y.O. equally contributed to this work. This work was supported by the National Research Foundation of Korea (NRF) grants RS-2021-017476, RS-2023-00218318, and RS-2023-00278985 (E.K., Y.O. and Y.B.) funded by the Ministry of Science and ICT (MSIT) of the Korea government, the NRF grant RS-2025-25438521 (Y.O.) funded by the Minstry of Education (MOE) of the Korea government, and individual KIAS Grant No.~QP10301 (E.K.) at the Korea Institute for Advanced Study. We also thank Youngkyoung Bae, Gleb Oshanin and Deepak Gupta for helpful discussions.

\begin{appendix}

\begin{widetext}

\section{Stochastic information flow in Langevin system} \label{sec:app_stoc_info_langevin}
In this section, we present an expression of stochastic information flow (SIF) in a bipartite multidimensional Langevin system and its relation with stochastic thermodynamics.

Consider a bipartite system described by two state vectors $\mathbf{x} (t) \in \mathbb{R} ^{N_x}$ and $\mathbf{y} (t) \in \mathbb{R} ^{N_y}$. Their time evolution follows
\begin{align}
    \dot{\mathbf{x}}(t) &= \mathbf{A}_x(\mathbf{x}(t),\mathbf{y}(t),t) 
        + \mathbb{B}_x(\mathbf{x}(t),\mathbf{y}(t),t)\,\bullet\, \boldsymbol{\xi}_x(t), \nonumber\\
    \dot{\mathbf{y}}(t) &= \mathbf{A}_y(\mathbf{x}(t),\mathbf{y}(t),t) 
        + \mathbb{B}_y(\mathbf{x}(t),\mathbf{y}(t),t)\,\bullet\, \boldsymbol{\xi}_y(t),
\end{align}
where $\bullet$ is a It\^o product, and $\boldsymbol\xi_x (t)$ and $\boldsymbol\xi_y (t)$ are two independent unit white Gaussian noises, $\mathbf{A}_x$ and $\mathbf{A}_y$ are drift field for each subsystem, and $\mathbb{B}_x$ and $\mathbb{B}_y$ are noise amplitude matrix for each subsystem. Since the noise injected to two dynamics is independent, the short-time propagator is factorized as 
\begin{align}
    \mathcal{P}(\mathbf{x}',\mathbf{y}',t{+}dt \mid \mathbf{x},\mathbf{y},t)
    = \mathcal{P}(\mathbf{x}',t{+}dt \mid\mathbf{x},\mathbf{y},t)\,
      \mathcal{P}(\mathbf{y}',t{+}dt\mid \mathbf{x},\mathbf{y},t),
\end{align}
with each subsystem short-time propagator satisfying (in the Stratonovich convention)
\begin{align}
    &\mathcal{P}(\mathbf{x}',t{+}dt|\mathbf{x},\mathbf{y},t)
     =\frac{1}{(4\pi dt)^{N_x/2}\, |\mathbb{D}_x|^{1/2}} \nonumber
    \\ \nonumber & \qquad \qquad \times \exp\!\Big[
        -\tfrac{dt}{4}\,
        \big(\dot{\mathbf{x}}_t - \mathbf{A}_x+ (\nabla_x ^\textsf{T} \mathbb{D}_x)^\textsf{T}\big)^{\mathsf T}
        \big(\mathbb{D}_x\big)^{-1}
        \big(\dot{\mathbf{x}}_t - \mathbf{A}_x + (\nabla_x ^\textsf{T}  \mathbb{D}_x)^\textsf{T}\big)
    \Big] 
    \\  & \qquad \qquad  \times 
    \exp\!\Big[
        -\tfrac{dt}{2}\,\nabla_x ^\textsf{T} \mathbf{A}_x
        + \tfrac{dt}{4}\sum_{ij}\partial_{x,i}\partial_{x,j} D_{x,ij}
    \Big],
\end{align}
where $\mathbb{D}_x=\tfrac{1}{2}\mathbb{B}_x\mathbb{B}_x^{\mathsf T}$ and $\nabla_x = (\partial_{x_1}, \partial_{x_2} , \cdots ,\partial_{x_{N_x}} )^\textsf{T}$ is the gradient operator acting on $\mathbf{x}$-subsystem. 
A similar expression holds for the $\mathbf{y}$-sector. With the short-time propagator, we can define entropy production rate as
\begin{align}
    \sigma_t = \lim_{dt \rightarrow 0 } \ln \frac{\mathcal{P} (\mathbf{x}_{t+dt},\mathbf{y}_{t+dt},t+dt;\mathbf{x}_t, \mathbf{y}_t,t)}{\mathcal{P} (\mathbf{x}_t, \mathbf{y}_t,t+dt;\mathbf{x}_{t+dt},\mathbf{y}_{t+dt},t)}
    \;,
\end{align}
where $\mathcal{P}(\mathbf{x}',\mathbf{y}',t+dt;\mathbf{x},\mathbf{y},t)$ is the joint probability of observing the system at the state $(\mathbf{x},\mathbf{y} )$ at time $t$ and $(\mathbf{x}',\mathbf{y}')$ at time $t+dt$.

The stochastic mutual information is given by
\begin{align}
    \mathcal{I}_t (\mathbf{x}_t, \mathbf{y} _t ) = \ln \frac{p_{\mathbf{x}_t,\mathbf{y}_t}(\mathbf{x}_t, \mathbf{y}_t)}{p_{\mathbf{x}_t}(\mathbf{x}_t)\,p_{\mathbf{y}_t}(\mathbf{y}_t)}
    \;
    \;.
\end{align}
Thus, we have
\begin{align}
    \dot{\mathcal{J}}_{\mathbf{x}_t} &= \partial_{t'} \mathcal{I}_t(\mathbf{x}_{t'}, \mathbf{y}_t)|_{t' =t} = \nabla_{\mathbf{x}} ^\textsf{T} \ln p_{\mathbf{y}_t \mid \mathbf{x}_t}(\mathbf{y}_t \mid \mathbf{x}) \big|_{\mathbf{x}=\mathbf{x}_t} \circ \dot{\mathbf{x}}_t \:, \nonumber \\ 
    \dot{\mathcal{J}}_{\mathbf{y}_t} &= \partial_{t'} \mathcal{I}_t(\mathbf{x}_t, \mathbf{y}_{t'})|_{t' =t} = \nabla_{\mathbf{y}} ^\textsf{T} \ln p_{\mathbf{x}_t \mid \mathbf{y}_t}(\mathbf{x}_t \mid \mathbf{y}) \big|_{\mathbf{y}=\mathbf{y}_t} \circ \dot{\mathbf{y}}_t 
    \;.
\end{align}
When the system is in the steady state, we have the relation $d_t \mathcal{I}_t (\mathbf{x}_t, \mathbf{y} _t ) = \dot{\mathcal{J}}_{\mathbf{x}_t} + \dot{\mathcal{J}}_{\mathbf{y}_t}$. In general, there is another contribution from $\partial_\tau \mathcal{I}_\tau (\mathbf{x}_t, \mathbf{y}_t ) |_{\tau = t}$. We have
\begin{align}
    \partial_{t} \mathcal{I}_t (\mathbf{x}_t, \mathbf{y}_t ) &= \frac{\partial_t p_{\mathbf{x}_t,\mathbf{y}_t}(\mathbf{x}_t,\mathbf{y}_t)}{p_{\mathbf{x}_t,\mathbf{y}_t}(\mathbf{x}_t,\mathbf{y}_t)} - \frac{\partial_t  p_{\mathbf{x}_t}(\mathbf{x}_t)}{p_{\mathbf{x}_t}(\mathbf{x}_t)}- \frac{\partial_t p_{\mathbf{y}_t}(\mathbf{y}_t)}{p_{\mathbf{y}_t}(\mathbf{y}_t)} \nonumber \\  
    & = \underbrace{\left[-\frac{\nabla_{\mathbf{x}}^\textsf{T} \mathbf{J}(\mathbf{x}_t, \mathbf{y} _t)}{p_{\mathbf{x}_t,\mathbf{y}_t}(\mathbf{x}_t,\mathbf{y}_t)} - \frac{\partial_t  p_{\mathbf{x}_t}(\mathbf{x}_t)}{p_{\mathbf{x}_t}(\mathbf{x}_t)} \right]}_{\equiv \dot{\mathcal{K}}_{\mathbf{x}_t}} +\underbrace{\left[-\frac{\nabla_{\mathbf{y}}^\textsf{T} \mathbf{J}(\mathbf{x}_t, \mathbf{y} _t)}{p_{\mathbf{x}_t,\mathbf{y}_t}(\mathbf{x}_t,\mathbf{y}_t)} - \frac{\partial_t  p_{\mathbf{y}_t}(\mathbf{y}_t)}{p_{\mathbf{y}_t}(\mathbf{y}_t)} \right]}_{\equiv \dot{\mathcal{K}}_{\mathbf{y}_t}}
    \;,
\end{align}
where last two lines define $\dot{\mathcal{K}}_{\mathbf{x}_t}$ and $\dot{\mathcal{K}}_{\mathbf{y}_t}$, namely the self-part of the SIF. Note that the ensemble average of the self-part vanishes for general time-dependent process, but the stochastic value itself does not vanish even for the system in the steady state. We define the total information flow $\dot{\mathcal{I}}_{\mathbf{x}_t} = \dot{\mathcal{J}}_{\mathbf{x}_t} + \dot{\mathcal{K}}_{\mathbf{x}_t}$ as the sum of flow and self part. Then, we obtain the decomposition $\partial_t \mathcal{I}_t = \dot{\mathcal{I}}_{\mathbf{x}_t} + \dot{\mathcal{I}}_{\mathbf{y}_t}$

Now, we relate the SIF with stochastic thermodynamics. Since the short-time propagator factorizes into two conditional probabilities, the total EP is 
\begin{align}
    \dot{s}_\text{tot} &= -d_t \ln p_{\mathbf{x}_t,\mathbf{y}_t}(\mathbf{x}_t ,\mathbf{y} _t ) \nonumber   \\ 
    &+ \lim_{dt\rightarrow 0} \frac{1}{dt}\left[  \ln \frac{\mathcal{P}(\mathbf{x}_{t+dt},t+dt|\mathbf{x}_t,\mathbf{y}_t,t)}{\mathcal{P}(\mathbf{x}_{t},t+dt|\mathbf{x}_{t+dt},\mathbf{y}_{t+dt},t)} + \ln \frac{\mathcal{P}(\mathbf{y}_{t+dt},t+dt|\mathbf{x}_t,\mathbf{y}_t,t)}{\mathcal{P}(\mathbf{y}_{t},t+dt|\mathbf{x}_{t+dt},\mathbf{y}_{t+dt},t)} \right]
    \;.
\end{align}
Note that 
\begin{align}
    \dot{s}_\text{sys} &= -d_t \ln p_{\mathbf{x}_t,\mathbf{y}_t}(\mathbf{x}_t ,\mathbf{y} _t ) \\ \nonumber 
    & = -d_t\mathcal{I}_t (\mathbf{x}_t ,\mathbf{y} _t) + d_t S_{\mathbf{x}_t}(\mathbf{x}_t) + d_t S_{\mathbf{y}_t}(\mathbf{y}_t)
    \;.
\end{align}
Thus, we have the decomposition
\begin{align}
    \dot{s}_{\mathrm{tot}}=\dot{\Sigma}_X - \dot{\mathcal{I}}_{\mathbf{x}_t} +\dot{\Sigma}_Y- \dot{\mathcal{I}}_{\mathbf{y}_t} 
    \;,
\end{align}
where $\dot{\Sigma}_X = \sigma_t ^{\text{env},X} + d_tS_{\mathbf{x}_t} (\mathbf{x}_t)$ is apparent entropy production for subsystem $X$, where
\begin{align}
    \sigma_t ^\text{env,X} &=\lim_{dt\rightarrow 0} \frac{1}{dt} \ln \frac{\mathcal{P}(\mathbf{x}_{t+dt},t+dt|\mathbf{x}_t,\mathbf{y}_t,t)}{\mathcal{P}(\mathbf{x}_{t},t+dt|\mathbf{x}_{t+dt},\mathbf{y}_{t+dt},t)}  \nonumber  \\ 
    &= \big[\mathbf{A}_x^{\mathrm{irr}} - (\nabla_x ^\textsf{T}  \mathbb{D}_x) ^\textsf{T}\big]^{\mathsf T}
         \mathbb{D}_x^{-1} \circ \big( \dot{\mathbf{x}}_t - \mathbf{A}_x^{\mathrm{rev}} \big)
      - \, \nabla_x ^\textsf{T} \mathbf{A}_x^{\mathrm{rev}}
    \;
\end{align}
is the environmental EP rate of the subsystem $X$, where $\mathbf{A}_x^\mathrm{rev}$ ($\mathbf{A}_x^\mathrm{irr}$) indicates the part of the drift $\mathbf{A}_x$ which changes (does not change) its sign under time reversal, therefore 
referred to as the reversible (irreversible) part of the drift.
Thus, we regard $\dot{\Sigma}_X - \dot{\mathcal{I}}_{\mathbf{x}_t}$ as the partial EP rate of the subsystem $X$.

Moreover, we can show that the partial EP rate satisfies the integral fluctuation theorem. To do so, let us define the adjoint (partial dual) by
\begin{align}
    \mathbf{A}_x^{a}(\mathbf{x},\mathbf{y},t)
    &= \mathbf{A}_x(\mathbf{x},\mathbf{y},t), \nonumber\\
    \mathbf{A}_y^{a}(\mathbf{x},\mathbf{y},t)
    &= -\mathbf{A}_y^{\mathrm{irr}}(\mathbf{x},\mathbf{y},t)
       + \mathbf{A}_y^{\mathrm{rev}}(\mathbf{x},\mathbf{y},t)
       + 2\, (\nabla_y ^\textsf{T} \mathbb{D}_y)^\textsf{T}
       + 2\, \mathbb{D}_y \nabla_y \ln p_{\mathbf{x}_t,\mathbf{y}_t}(\boldsymbol{\epsilon}\mathbf{x},\boldsymbol{\epsilon}\mathbf{y})\:,
\end{align}
where the operator $\boldsymbol{\epsilon}$ flips the sign of the odd-parity variables such as momentum, and keeps that of the even-parity variables such as position.
Then, one can verify that
\begin{align}
    \lim_{dt\rightarrow 0} \frac{1}{dt}\ln \frac{\mathcal{P}(\mathbf{x}',\mathbf{y}',t{+}dt |\mathbf{x},\mathbf{y},t)}
             {\mathcal{P}^{a}(\mathbf{x},\mathbf{y},t{+}dt | \mathbf{x}',\mathbf{y}',t)}
    = \dot{\Sigma}_X - \dot{\mathcal{I}}_{\mathbf{x}_t} \equiv \dot{\tilde{s}}_X,
\end{align}
so that the integral fluctuation theorem (IFT) holds:
\begin{align}
    \left\langle e^{-[\Sigma _X (\tau) - \mathcal{I}_{\mathbf{x}} (\tau)]} \right\rangle = 1
    \;,
\end{align}
where $\Sigma_\tau ^X  = \int_0 ^\tau dt \, \dot{\Sigma}_t ^X$ and $\mathcal{I}_{\mathbf{x}} (\tau) = \int_0 ^\tau dt \,\dot{\mathcal{I}}_{\mathbf{x}_t}$ are the quantities accumulated over time.

\section{Stochastic information flow in Markov jump system} \label{sec:app_stoc_info_MJ}
In this section, we present an expression of SIF in bipartite discrete Markov jump system and its relation with stochastic thermodynamics. Let the system have two degrees of freedom, $X=x$ and $Y = y$. 
In a bipartite system, only one degree of freedom is allowed to change at a time. 
Then, $W_{xx',y}$ ($W_{x,yy'}$) denotes the transition rate from $X=x'$ ($y'$) to $x$ ($y$) while $y$ ($x$) is fixed. 
The master equation is
\begin{align}
    \partial_t p_{X_t,Y_t}(x,y)
    = \sum_{x \neq x'} j_{t}(x,x';y) + \sum_{y \neq y'} j_{t}(x;y,y')
    \;,
\end{align}
where $j_{t}(x,x';y)  = W_{xx',y} p_{X_t,Y_t}(x',y) - W_{x'x,y} p_{X_t,Y_t}(x,y)$ is the net probability current 
from state $x'$ to $x$ while $y$ is fixed 
(and $j_{t}(x;y,y')$ is defined analogously).

We begin by writing the stochastic mutual information as
\begin{align} \label{eq:MJ_SMI}
    \mathcal{I}_t (X_t, Y_t ) = \sum_{x,y} \ln \frac{p_{X_t,Y_t} (x,y)}{p_{X_t} (x) \, p_{Y_t}(y)} \eta _{t} (x,y)
    \;,
\end{align}
where $p_{X_t}(x) = \sum_{y} p_{X_t,Y_t}(x,y)$, $p_{Y_t}(y) = \sum_x p_{X_t,Y_t}(x,y)$, and  $\eta_{t}(x,y)\equiv \delta_{X_t,x}\delta_{Y_t,y}$ is the state indicator. Note that
\begin{align}
    \lim_{dt \rightarrow 0} \frac{\delta_{X_{t+dt},x} \delta _{Y_t,y}- \delta_{X_t, x}\delta _{Y_t,y}}{dt} = \sum_{x' (\neq x)}[\dot{N}_{t} (x,x'; y) - \dot{N}_{t} (x',x; y)]
    \;,
\end{align}
where $N_{t} (x,x';y)$ increases by $1$ whenever there is a transition from $x'$ to $x$ while the state $Y=y$ is fixed. Then, we have
\begin{align}
    \dot{\mathcal{J}}_{X_t} &= \partial_\tau \mathcal{I}_t(X_{\tau}, Y_t)|_{\tau =t} \nonumber \\ 
    &= \sum_{x \neq x',y}  \ln \frac{p_{X_t,Y_t} (x,y)}{p_{X_t} (x) \, p_{Y_t}(y)} [\dot{N}_{t} (x,x';y) - \dot{N}_{t} (x',x;y)]\nonumber   \\ 
    & = \sum_{x \neq x',y} \ln \frac{p_{Y_t\mid X_t}(y\mid x)}{p_{Y_t\mid X_t}(y\mid x')} \dot{N}_{t} (x,x';y )
    \;,
\end{align}
where $p_{Y_t\mid X_t} (y\mid x)$ is the conditional probability of observing $Y_t=y$ given $X_t =x $. Likewise, the SIF rate flowing into $Y$ is given by
\begin{align}
    \dot{\mathcal{J}}_{Y_t} = \sum_{y\neq y', x} \ln \frac{p_{X_t\mid Y_t} (x\mid y)}{p_{X_t\mid Y_t}(x\mid y')} \dot{N}_{t}(x;y,y')
    \;,
\end{align}
where $N_{t} (x;y,y')$ increases by $1$ whenever there is a transition from $y'$ to $y$ while the state $X=x$ is fixed.

When the system is in the steady state, we have the relation $d_t \mathcal{I}_t (X_t ,Y_t) = \dot{\mathcal{J}}_{X_t} + \dot{\mathcal{J}}_{Y_t}$. In general, there is another contribution from $\partial_\tau \mathcal{I}_{\tau} (X_t, Y_t) |_{\tau =t}$. Inside the sum in the RHS of Eq.~\eqref{eq:MJ_SMI}, the explicit dependence through $t$ only resides in $ \ln \frac{p_{X_t,Y_t} (x,y)}{p_{X_t} (x) \, p_{Y_t}(y)} $. Thus, we have
\begin{align}
    \partial_\tau \mathcal{I}_{\tau} (X_t, Y_t) |_{\tau =t} & = \frac{\partial _t p_{X_t,Y_t}(x,y)}{p_{X_t,Y_t} (x,y)} - \frac{\partial_t p_{X_t} (x)}{p_{X_t} (x)} - \frac{\partial_t p_{Y_t} (y)}{p_{Y_t} (y)} \nonumber \\ \nonumber 
    &= \underbrace{\sum_{x,y} \eta_t (x,y) \left[\frac{\sum_{x (\neq x')} j_{t}(x,x';y) }{p_{X_t,Y_t} (x,y)} - \frac{\partial_t p_{X_t} (x)}{p_{X_t} (x)}\right]}_{\equiv\dot{\mathcal{K}}_{X_t}} \\  & \quad  + \underbrace{\sum_{x,y} \eta_{t} (x,y) \left[\frac{\sum_{y'( \neq y')} j_{t}(x;y,y') }{p_{X_t,Y_t} (x,y)} - \frac{\partial_t p_{Y_t} (y)}{p_{Y_t} (y)}\right]}_{\equiv\dot{\mathcal{K}}_{Y_t}}
    \;,
\end{align}
where last two lines define $\dot{\mathcal{K}}_{X_t}$ and $\dot{\mathcal{K}}_{Y_t}$, namely the self-part of the SIF. Note that the ensemble average of the self-part vanishes for general time-dependent process, but the stochastic value itself does not vanish even for the system in the steady state. 

To relate the SIF with stochastic thermodynamics, we recall that the stochastic entropy production rate in this system is given by
\begin{align}
    \sigma _t &= -d_t \ln p_{X_t,Y_t}  (X_t, Y_t)  \nonumber \\ \nonumber 
    & \quad +  \sum_{x\neq x',y} \ln\frac{W_{xx',y}}{W_{x'x,y}} \dot{N}_t (x,x';y) + \sum_{x,y\neq y'} \frac{W_{x,yy'}}{W_{x,y'y}} \dot{N}_t (x,x';y) \\ \nonumber 
    & = \underbrace{\sum_{x\neq x',y} \ln\frac{W_{xx',y}p_{X_t,Y_t}(x',y)}{W_{x'x,y}p_{X_t,Y_t}(x,y)}  - \sum_{x,y} \eta_t (x,y)\frac{\sum_{x (\neq x')} j_{t}(x,x';y) }{p_{X_t,Y_t} (x,y)}}_{\equiv \sigma _t ^X} \\  
    & \quad + \underbrace{ \sum_{x,y\neq y'} \ln  \frac{W_{x,yy'}p_{X_t,Y_t} (x,y')}{W_{x,y'y}p_{X_t,Y_t} (x,y)} \dot{N}_t (x,x';y) - \sum_{x,y} \eta_t (x,y) \frac{\sum_{y'( \neq y')} j_{t}(x;y,y') }{p_{X_t,Y_t} (x,y)}}_{\equiv \sigma _t ^Y}
    \;,
\end{align}
where we use $\ln p_{X_t,Y_t} (X_t ,Y_t) = \sum_{x,y} \eta_t (x,y) \ln p_{X_t,Y_t} (x,y)$ in the second equality. Then, we have
\begin{align}
    \sigma_t ^X + (\dot{\mathcal{J}}_{X_t} + \dot{\mathcal{K}}_{X_t})&=\sum_{x \neq x',y} \ln \frac{W_{xx',y} p _{X_t} (x')}{W_{x'x,y}p_{X_t} (x)} \dot{N}_t (x,x';y) - \sum_{x,y} \eta_t (x,y)  \frac{\partial_t p_{X_t} (x)}{p_{X_t}(x)} \nonumber  \\ 
    & = \sigma ^{\text{env},X}_t + d_t S_{X_t} (X_t)
    \;.
\end{align}
Thus, by defining the total information flow $\dot{\mathcal{I}}_{X_t} =\dot{\mathcal{J}}_{X_t} +\dot{\mathcal{K}}_{X_t}  $ and apparent entropy production $\dot{\Sigma}_t ^X$, we have the relation $\sigma _t ^X = \dot{\Sigma}_t ^X - \dot{\mathcal{I}}_{X_t}$. 

Moreover, we can construct the IFT, which is done similarly in Ref.~\cite{Shiraishi2015fluctuation,Bisker2017hierarchical}. We first present the probability of observing the trajectory $\Gamma = \{ X_t , Y_t\}_{t=0} ^\tau$:
\begin{align}
    \mathcal{P}[\Gamma] &= p_{X_0,Y_0} (X_0, Y_0) \exp \left[ \int_0 ^\tau dt \sum_{x,y} \eta_t (x,y) (W_{xx,y}+W_{x,yy}) \right]  \nonumber \\  
    & \quad \times \exp \left\{\int_0 ^\tau dt \left[\sum_{x\neq x', y} \ln W_{xx',y} \dot{N}_{t} (x,x';y) + \sum_{x,y\neq y'} \ln W_{x,yy'} \dot{N}_{t} (x;y,y') \right]\right\}
    \;.
\end{align}
Next, let us define an auxiliary process with rates~\cite{Shiraishi2015fluctuation}
\begin{align}
    W^a_{xx',y} = W_{xx',y},
    \qquad
    W^a_{x,yy'} = W_{x,y'y}\, \frac{p_{X_t,Y_t}(x,y)}{p_{X_t,Y_t}(x,y')}
    \;.
\end{align}
The escape rate through $x$-jumps is unchanged, since
\begin{align}
    W^a_{xx,y}
    = -\sum_{x' \neq x} W^a_{x'x,y}
    = -\sum_{x' \neq x} W_{x'x,y}
    = W_{xx,y}
    \;,
\end{align}
while the escape rate through $y$-jumps becomes
\begin{align}
    W^a_{x,yy}
    = -\sum_{y' \neq y} W^a_{x,y'y}
    = \frac{\sum_{y' \neq y} W_{x,yy'} p_{X_t,Y_t}(x,y')}{p_{X_t,Y_t}(x,y)}
    \;.
\end{align}
The log-ratio between the path probabilities of the original and reverse auxiliary processes is
\begin{align}
    \ln \frac{\mathcal{P}[\Gamma]}{\tilde{\mathcal{P}}^\text{A}[\tilde{\Gamma}]}
    &= \ln \frac{p_{X_0,Y_0}(X_0,Y_0)}{p_{X_\tau, Y_\tau} (X_\tau, Y_\tau)}
       + \int_0^\tau dt 
       \bigg[
           \sum_{x \neq x',y} \ln \frac{W_{xx',y}}{W_{x'x,y}}\, \dot{N}_{t}(x,x';y)
           + \sum_{x,y \neq y'} \ln \frac{p_{X_t,Y_t}(x,y)}{p_{X_t,Y_t}(x,y')}\, \dot{N}_{t}(x;y,y')
       \bigg] \nonumber\\
    &\quad \quad \quad \quad \quad - 
    \int_0^\tau dt \sum_{x,y} \eta_{t}(x,y)
       \left[
           W_{x,yy} 
           - \frac{\sum_{y' \neq y} W_{x,yy'} p_{X_t,Y_t}(x,y')}{p_{X_t,Y_t}(x,y)}
       \right] \nonumber\\
    &= \int_0^\tau dt \sum_{x \neq x',y} 
       \ln \frac{W_{xx',y} p_{X_t,Y_t}(x',y)}{W_{x'x,y} p_{X_t,Y_t}(x,y)}\, \dot{N}_{t}(x,x';y) \nonumber
       \\ 
       & \quad \quad \quad \quad \quad+ 
       \int_0^\tau dt \sum_{x,y} \eta_{t}(x,y)
       \underbrace{\left[
           \frac{\sum_{y' (\neq y)} j_{t}(x;y,y')}{p_{X_t,Y_t}(x,y)} 
           - \partial_t \ln p_{X_t,Y_t}(x,y)
       \right]}_{= -\frac{\sum_{x' (\neq x)} j_{t}(x,x';y)}{p_{X_t,Y_t}(x,y)}} \nonumber\\
    &= \int _0 ^\tau dt \, \sigma_t ^X
    \;.
\end{align}
Thus, we obtain the IFT
\begin{align}
    \left\langle e^{-[\Sigma _\tau ^X -  \mathcal{I}_{X} (\tau)]} \right\rangle = 1
    \;,
\end{align}
where $\Sigma_\tau ^X  = \int_0 ^\tau dt \dot{\Sigma}_t ^X$ and $\mathcal{I}_X (\tau) = \int_0 ^\tau dt\, \dot{\mathcal{I}}_{X_t}$. The subsystem second law $\langle \Sigma _\tau ^X \rangle \ge \langle \mathcal{I}_X (\tau ) \rangle $ naturally follows from the IFT. While the self-part $\dot{\mathcal{K}}_{X_t}$ plays no role in the subsystem second law, but the fluctuation of the term affects the higher order moments of $\Sigma _\tau ^X$. 

\begin{figure}
	\includegraphics[width=\textwidth]{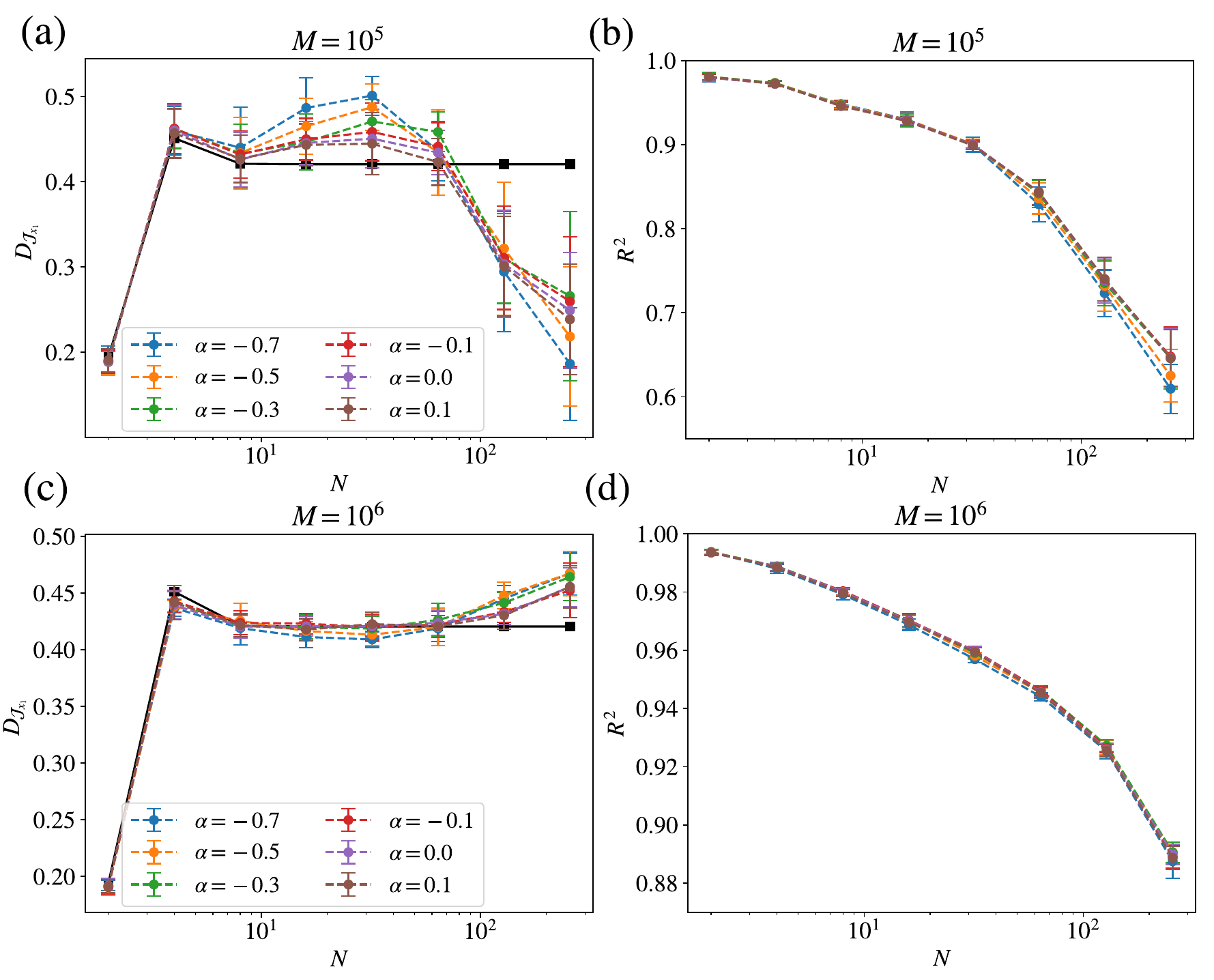}
    \caption{\label{fig:effect_alpha} 
    (a,c) The scaled variance ($D_{ \mathcal{J}_{x_1}}$) and (b, d) the coefficient of determination ($R^2$) between $\Delta\mathcal{J}_{x_1}(\tau = 1)$ and $\Delta \hat{\mathcal{J}}_{x_1 }(\tau = 1)$, estimated by NESIF with different $\alpha$. Dataset size is given by $M=10^5$ in (a,b) and $M=10^6$ in (c,d). 
    }
\end{figure}

\section{Estimating stochastic information flow with $\alpha$-MINE} \label{sec:app_alpha_mine}

Here, we present a generalized version of the MINE, which we call $\alpha$-MINE. We define $\alpha$-divergence between the joint distribution $p_{XY} (x,y)$ and the product of marginals $p_X (x) p_Y (y)$:
\begin{align} \label{eq:alpha_div_def}
D_\alpha [ p_{XY} : p_X p_Y ] \equiv \frac{1}{\alpha (1+\alpha)} \left\langle \left[ \frac{p_{XY}(x,y)}{p_X (x) p_Y (y)} \right]^\alpha -1\right\rangle_{p_{XY}}
\;,
\end{align}
where $\alpha$ is a real number. For the case of $\alpha = 0$ and $\alpha = -1$, we understand Eq.~\eqref{eq:alpha_div_def} by taking the appropriate limit. This is generalization of the mutual information which can be recovered in the $\alpha \rightarrow 0$ limit. The variational representation of $\alpha$-divergence is given by
\begin{align} \label{eq:alpha_div_var}
D_{\alpha} [p_{XY}: p_X p_Y] &\ge \left\langle h_\theta (x,y;\alpha)\right\rangle_{p_{XY}}- \left\langle h_\theta (x,y;\alpha) \right\rangle_{p_X p_Y} \\ \nonumber &\equiv \mathcal{C}_{\alpha}(\theta)
\;,
\end{align}
where $h_\theta (x,y;\alpha) = (e^{\alpha \mathcal{I}_\theta (x,y) }-1)/\alpha$ and $\mathcal{I}_\theta (x,y)$ is the estimator of the SMI with parameters of the neural network given by $\theta$. 

The derivation of Eq.~\eqref{eq:alpha_div_var} follows from Ref.~\cite{Kwon2024alpha}, and we repeat the derivation for the completeness. Let $f(u)$ be the following function:
\begin{align} \label{eq:f_alpha}
    f_\alpha (u) = 
    \begin{cases}
    \frac{u^{1+\alpha}-(1+\alpha)u+\alpha}{\alpha (1+\alpha)} & \text{for $\alpha \neq 0,\,-1$,} \\
    u\log u & \text{for $\alpha = 0$,} \\
    \log u + 1 - u & \text{for $\alpha = -1$.}
    \end{cases}
    \;.
\end{align}
Then, it can be easily checked that $f(u)$ is convex. Then, we can verify that the inequality 
\begin{align} \label{eq:fineq}
pf'(u)-q \left[uf'(u) - f(u) \right] \le qf(p/q)
\;
\end{align}
holds, since the LHS has a unique local maximum at $u=p/q$. 
Then, we put $p=p_{XY} (x,y)$, $q=p_X(x) p_Y (y)$, $u=e^{i_\theta (x,y)}$, and integrate with respect to $x$ and $y$ to get Eq.~\eqref{eq:alpha_div_var}. The equality holds if and only if $i_\theta (x,y) = \ln \frac{p_{XY} (x,y)}{p_X (x) p_Y (y)} =i(x,y)$, \emph{i.e.} when the estimator matches with the PMI. Then, by maximizing the utility function $\mathcal{C}_\alpha (\theta )$, we expect that the $\theta$ converges to $\theta ^*$, where $i_{\theta^*}(x,y)$ is the best estimator of $i(x,y)$. 

After training $\alpha$-MINE, the SIF is similarly estimated by
\begin{align} \label{eq:SIF_pred}
    \hat{\dot{\mathcal{J}}}_{X_t} = \frac{i_{\theta ^*} (X_{t+\Delta t}, Y_t ) - i_{\theta ^*}(X_t, Y_t)}{\Delta t}
    \;.
\end{align}
We investigate the effect of $\alpha$ in estimating the SIF in Fig.~\ref{fig:effect_alpha}, using the $N$-bead model introduced in the main text. For the dataset size $M=10^5$, we observe that the scaled variance ($D_{ \mathcal{J}_{x_1}}$) is predicted well for $N\le8$, and deviates from the exact value as $N$ is increased. For moderate value of $16\le N \le 64$, the NESIF overestimates the scaled variance, while for higher values of $N$, it underestimates the scaled variance. We also observe that the coefficient of determination ($R^2$) is decreasing as we increase $N$. For different $\alpha$ values plotted, $\alpha =0.1$ seems to work best for $N\le 64$, but the data from $\alpha =-0.1$ and $\alpha=0$ (which corresponds to the original NESIF used in the main text) also lie within the error bars. The results are in contrast with Ref.~\cite{Kwon2024alpha}, where we reported that $\alpha =-1/2$ works best when estimating stochastic entropy production using variational representation of $\alpha$-divergence. For $M=10^6$, the performance of NESIF gets much better than $M=10^5$, and the effect of $\alpha$ is decreased. At this level, we could not conclude which $\alpha$ works best, and we left this point as a future work.

\section{Cooperative dynamics of Kuramoto oscillators without external driving} \label{sec:app_Kuramoto}

\begin{figure}
	\includegraphics[width=\textwidth]{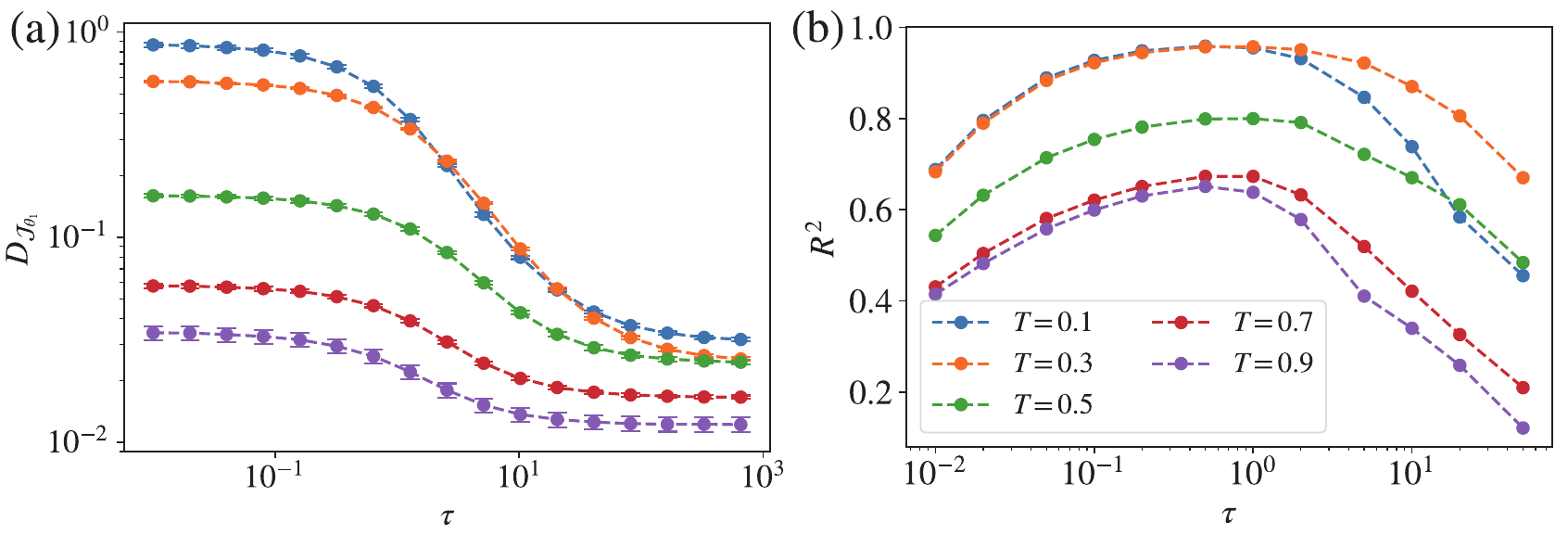}
    \caption{\label{fig:sub_kuramoto} 
     SIF statistics of $32$ noisy Kuramoto oscillators with $K = 1$ simulated and recorded using the time step size $\Delta t=0.01$. Colors indicate temperatures. (a) Scaled variance $D_{\hat{\mathcal{J}}_{\theta_1}}$ as a function of the observation period $\tau$. (b) The coefficient of determination $R^2$ between the change in cosine similarity $\Delta C (\tau)$ and the time-integrated SIF $\hat{\mathcal{J}}_{\theta_1} (\tau)$. The lines are to guide the eye.
    }
\end{figure}

In this section, we provide a detailed description of cooperative dynamics of Kuramoto model in the simplest case of no external driving: $g(\omega) = \delta(\omega)$. As was done for the $N$-bead model, we measure the time-integrated SIF into a single oscillator from the rest of the system. While the symmetry forces the mean SIF to vanish, we observe that the scaled variance $D_{\mathcal{J}_{\theta_1}}$ decreases monotonically with temperature $T$, as shown in Fig.~\ref{fig:sub_kuramoto}(a). This indicates that oscillators exchange information more vigorously when $T$ is lower, so that they synchronize more. In addition, $D_{\mathcal{J}_{\theta_1}}$ decreases with the observation period $\tau$, with the decrease accelerating between $\tau \simeq 10^{-1}$ and $\tau \simeq 10^0$. This suggests the existence of a characteristic time scale $\tau^*$, which corresponds to the average time it takes for an oscillator to leave a cluster of synchronized oscillators. When $\tau \ll \tau^*$, every oscillator tends to move within a cluster of synchronized oscillators throughout the observation time, vigorously exchanging information with them. When $\tau \gg \tau^*$, an oscillator leaves and joins the cluster intermittently, exchanging much less information with others during the periods of separation. To corroborate this conjecture, in Fig.~\ref{fig:sub_kuramoto}(b), we plot the coefficient of determination $R^2$ between $ \mathcal{J}_{\theta_1} (\tau)$ and $\Delta C(\tau;t) \equiv C(t+\tau) - C(t)$, where $C(t) \equiv \cos (\theta _ 1(t) - \bar{\theta}_{-1} (t))$ is the cosine similarity between the phase $\theta_1(t)$ of one oscillator and the average phase $\bar{\theta}_{-1} (t)$ of the rest defined via $\bar{\theta}_{-1} (t) \equiv \arg \left[\frac{1}{N-1} \sum_{n=2}^N \mathrm{e}^{i\theta_n(t)}\right]$.

\end{widetext}

\end{appendix}

\bibliography{bibliography}

\end{document}